\documentstyle[12pt,aasms4]{article}

\lefthead{Garz\'on et al.}
\righthead{A SFR in the Galactic plane at $\ell=27$\arcdeg}

\begin{document}

\title{A major star formation region in the receding tip of the stellar
Galactic bar}

\author{F. Garz\'on, M. L\'opez--Corredoira, P. Hammersley, T.J.
Mahoney, X. Calbet and J.E. Beckman}
\affil{Instituto de Astrof\'{\i}sica de Canarias, E--38200 La Laguna
(Tenerife) SPAIN}
\authoraddr{E--38200 La Laguna (Tenerife) SPAIN}

\begin{abstract}

We present an analysis of the optical spectroscopy of 58 stars in the
Galactic plane at $\ell=27$\arcdeg, where a prominent excess in the
flux distribution and star counts have been observed in several
spectral regions, in particular in the Two Micron Galactic Survey
(TMGS) catalog. The sources were selected from the TMGS, to have a $K$
magnitude brighter than +5 mag and be within 2 degrees of the Galactic
plane. More than 60\% of the spectra correspond to stars of luminosity
class I, and a significant proportion of the remainder are very late
giants which would also be fast evolving. This very high concentration
of young sources points to the existence of a major star formation
region in the Galactic plane, located just inside the assumed origin of
the Scutum spiral arm. Such regions can form due to the concentrations of
shocked gas where a galactic bar meets a spiral arm, as is observed at
the ends of the bars of face-on external galaxies. Thus, the presence
of a massive star formation region is very strong supporting evidence
for the presence of a bar in our Galaxy.

\end{abstract}

\keywords{stars: formation -- Galaxy: stellar contents -- Galaxy:
structure.}

\section {Introduction}

The barred nature of the Milky Way was first suggested by de
Vaucouleurs (1964) in an attempt to explain the non-circular gas
orbits. Since then, many types of observational supporting evidence
have been accumulated. These comprise modern star-count data and
surface photometry at different wavelengths, which suggest axial
asymmetry of the internal bulge, stellar population studies in Baade's
window, microlensing, and detailed analysis of the internal motions of
the gas. Many review papers are cited by Blitz (1996).

In our analysis of the Two Micron Galactic Survey (TMGS) database
(Garz\'on et al. 1993) we described evidence in favor of a central
Galactic bar. In this paper, we present the results of a spectroscopic
follow-up to the TMGS project, aimed at identifying the population
content of the Galactic plane in selected regions. Here we concentrate
on the $\ell=27$\arcdeg\  area, where excesses in the flux distribution
are very clearly observed in scans across the plane in the TMGS
database and have been reported by previous authors at different
wavelengths (Viallefond et al. 1980), to explore the hypothesis that we
are looking at a star-forming region of particular strength.

\section{TMGS star counts in the Galactic plane at $\ell=27$\arcdeg.}

The TMGS catalog is composed of a series of slices at constant
declination across the Galactic plane, each scan covering up to 30
degrees of latitude and a variable amount in longitude, ranging from
0.1 to 2.5 degrees (for details see Garz\'on et al. 1993). Hence the
areas we are referring to here  cover an area on the sky of roughly one
degree in $b$, centered on the Galactic equator, by the amount in
$\ell$ scanned by the TMGS at the given location.

Using TMGS counts Hammersley et al. (1994, hereafter  HGMC) showed
that at $\ell=21$\arcdeg\  and $27$\arcdeg there is a concentration of
luminous stars within about a degree of the plane, which, when the
counts are plotted against latitude, form a spiked distribution on top
of that expected for the disk and arms. Hence these latter two Galactic
components cannot account for the observed stellar distribution.

HGMC argued that the  spikes detected in the TMGS star counts and in
the DIRBE surface photometry are both due to the interaction between
the  ends of the bar and the Galactic disk, which gives rise to star
formation regions. Calbet et al. (1996) also pointed out that the
infrared star-count distribution in the central regions can be
explained in terms of a dust lane leading a Galactic bar at negative
longitudes.

A clear excess in the star distribution is observed in the Galactic
plane at $\ell=27$\arcdeg\  when compared with neighboring areas;
namely, those at $\ell=33$\arcdeg, $\ell=21$\arcdeg\,  and
$\ell=16$\arcdeg\ (HGMC). HGMC examined several possible scenarios to
account for this excess, which is also noticeable in other types
surveys (Viallefond et al. 1980; Hayakawa et al. 1981; Kawara et al.
1982; Kent et al. 1991).  This spike is especially noticeable when  the
sample is limited to apparent magnitudes brighter than $m_K=5$ mag. We
concluded that the most likely explanation is the presence of a giant
star formation region, which is most probably associated with the
receding tip of the Galactic bar.


Mikami et al. (1982) studied this region using an objective-prism plate
survey, and suggested that clustering of M supergiants must be the
cause of the peak. We found their suggestion in agreement with our
conclusion and, with this in mind, started an observational program
aimed at spectroscopically classifying  selected sources from the TMGS
database, particularly those responsible for the spike at $\ell
=27$\arcdeg.

\section{Observations and analysis}

The observations were carried out at the Roque de los Muchachos
Observatory on La Palma (Canary Islands, Spain) with the 2.5-m Isaac
Newton Telescope using the Intermediate Dispersion Spectrograph (IDS),
with the AgRed collimator and the R600IR grating (centered at
8500--8600 \AA) and TEK3 chip CCD. The spectra cover the region from
7750 to 9400 \AA\ and the spectral resolution attained with two
selected instrumental configurations was $1.7$ \AA\ pix$^{-1}$.

We ran three campaigns in the summers of 1995 and 1996, taking some 70
spectra in total, 60 of which were identified as visible counterparts
of TMGS sources. The targets were selected from the TMGS database in
the Scutum region ($\ell=27$\arcdeg, $b=0$\arcdeg) using $m_K<5$ mag as
the selection criterion. After standard reduction, we examined the
spectra for the IR Ca {\sc ii} triplet at 8498.02 \AA, 8542.09 \AA\ and
8662.14 \AA,  present in stars of spectral types later than F5. In
earlier types, the Paschen hydrogen lines severely contaminate the
spectral region of interest, making the Ca {\sc ii} triplet difficult
to measure. For spectral types later than M4--5, TiO absortion bands
mask the Ca {\sc ii} triplet almost completely. Jones et al. (1984) and
D\'{\i}az et al. (1989) have calibrated the relationship between the
equivalent width (EW) of the Ca {\sc ii} triplet and the luminosity
class empirically for spectral types ranging from F5 to M3.

In Table \ref{Tabla} we give the coordinates of all the stars in our
sample  from the TMGS database. Investigators who may wish
to use them should be aware that due to the intrinsic unaccuracies of the
TMGS coordinates, which can be roughly estimated in 5\arcsec\  to
10\arcsec,  and the extreme crowdeness of the field there is often more
than one visible candidate for the IR source.

\placetable{Tabla}

These authors found some dependence of  EW on both metallicity and
temperature, but these were much weaker than the dependence on surface
gravity. We have followed their results and adopted their criteria in
assigning luminosity classes from the measured EWs. In practice, we
classify a star as a supergiant (SG) by means of the EW of the two
stronger triplet lines only in order to minimize the errors. The source
is assumed to be a SG if the sum of the EWs of these two lines is $>9$
\AA, as in D\'{\i}az et al. (1989). The definition of the working
continuum from which the EW is measured also followed that of D\'{\i}az
et al. (1989). 38 stars in the sample were not contaminated with the
TiO band, and we used the above method to get the EW. The remaining 22
stars belong to later spectral types, and the presence of TiO bands
affects the triplet lines. For these objects we have evolved an
empirical method which permits the measurement of EW in cases where the
Ca {\sc ii} lines were not completely masked by the TiO band. This
method uses the depth of the lines instead of the EW. First, we
calibrated the relationship between line depth and the aggregated EW in
the 38 stars where both quantities were measurable. We then used this
relation to predict the EW from the measured line depth, where the Ca
{\sc ii} lines are partially masked by the TiO band. Even with this
technique, we had to reject two stars of the sample since the TiO band
gave an unacceptable blend, so we finished  with 58 stars that could be
used.

The final results of the luminosity classification are shown in Fig.
\ref{Fig:histo}, in the form of a histogram of EW frequencies. Most
noteworthy is the ratio of SGs (those with EW $> 9 \AA$) to the total
number, well in excess of 50\%, with a high degree of confidence. The
number of SGs is in fact 36 out of 58 (62\%). This is strong evidence
for the presence of a cluster of SGs, associated with a star-forming
region. According to the model of Wainscoat et al. (1992), the disk
and spiral arms can account for a maximum of  only 20\% of the SGs in
this region. Furthermore, this model also predicts that there should
be about 20 giants per square degree in the area, which is in
approximate agreement with the number of giants found in this work. The
remaining stars are either giants or dwarfs. For these a precise
segregation is not quite so straightforward since in these classes the
effect of metallicity in the relationship between EW and $\log g$ is
stronger. Following the above-mentioned criteria,  however, we can
classify as giants those with EWs  between 6 and 9 \AA, and as dwarfs
those with EW $< 6$ \AA. According to this criterion, there are no
dwarfs in the sample, as expected from the $K$ magnitude limit used for
the selection of the sample.


\subsection{Spectral classification}

We have also performed a spectral classification by comparing our
spectra with those of standard stars taken from the literature
(Barbieri et al. 1981; Schulte-Ladbeck 1988; Bessel 1991; Torres-Dodgen
\& Weaver 1993). In Fig. \ref{Fig:histo_tipo} we show the frequency
distribution of the spectral classes. As expected from objects selected
on the basis of their $K$ magnitudes, most of the stars are very red.

We also show in Fig. \ref{Fig:histo_tipo_sg} a histogram of the
frequency of different spectral types for the SGs only. For the SGs K
is the most frequent spectral type, although our method of predicting
the EWs where the TiO band affects the Ca {\sc ii} lines tends to
underestimate the EWs, thereby reducing the apparent fraction of SGs in
the coolest (M) class.


\section{Star formation region}

According to Bica et al. (1990a, 1990b) the red supergiant phase in a
cluster is reached at an age of $\sim 10$ Myr. This implies that we are
looking at a region in which the star formation is of recent origin,
and that the star formation has taken place in the area observed, since
such a short time does not permit the stars to move very far  from
their birthplace. This means that we can estimate the distance to the
star forming region from the TMGS apparent $K$ magnitudes by using the
absolute $K$ magnitudes associated with known spectral types and
luminosity classes, which can be found in the literature (Johnson 1966;
Lee 1970; Blaauw 1973; Ishida \& Mikami 1978). Since these sources are buried deep in the Galactic plane (we estimate their visible magnitudes to be about 16 mag), the probability of finding counterparts in existing visible catalogs is virtually nil. We adopted the extinction
model of Wainscoat et al. (1992).


The result, in terms of the distance distribution, is plotted in Fig.
\ref{Fig:histo_dist}. This histogram shows two maxima, one at
distances of between 2 and 3 kpc and the other peaking at around 6 kpc.  The
first peak can be attributed to young-disk supergiants along the line of sight. The second peak is more spread out, but this dispersion can
be explained by the uncertainty in the distance determination (due to
errors in both the TMGS photometry and the assumed absolute
magnitudes), which increases with distance. This peak, then, is formed
by stars at distances ranging from 5 to 8 kpc;  24 of the 36 SGs belong
to this peak. This SG concentration cannot be explained in terms of
disk and/or bulge population, since these are mainly formed by old
stars, so it must be attributable to some other component.

Since this region is not prominent among the main H {\sc ii} radio
sources in the Galactic plane (Georgelin \& Georgelin 1976), it must be
embedded in another feature, such as an arm,  a ring or  a bar. Several
authors have used Galactic models which include ring components to
explain the excess flux observed in these regions (Mikami et al. 1982;
Ruelas-Mayorga 1991; Kent et al. 1991). However, HGMC argued that if a
ring is to be the feature where this star-forming region is located,
then the ring has to be highly non-circular and discontinuous, as the
observed peaks in star counts are not symmetrically distributed in
longitude with respect to the Galactic center.

The arms can be quickly discarded since their tangential cuts are not
in the direction of this region. From a number of morphological and
modeling arguments discussed in HGMC, we have reason to think that
neither a spiral arm nor a ring can be the morphological structure
within which this region is situated. A ring can also be excluded with
fair probability since it should be prominent in other TMGS regions
closer to the center than $\ell$=27\arcdeg, which is not the case. This
argument is of course not as powerful as that for excluding arm
sources.

\section{Conclusion}

We have found a star formation region located a $\ell=27$\arcdeg\  in
the Galactic plane. This region extends, presumably, to at least
$\ell=21$\arcdeg, as can be deduced from the star distribution in the
TMGS. The most likely explanation is that the Milky Way is a barred
galaxy, and that this star-forming region is the result of the
interaction between the suggested bar and the Scutum spiral arm. HGMC
inferred a maximum position angle for the bar of 75\arcdeg by
considering that the ends of the bar are located at $\ell=27$\arcdeg\
and $-22$\arcdeg. This geometry is compatible with the range of
distances that we have obtained for the star-forming region.

\acknowledgements

The Isaac Newton Telescope is operated on the island of La Palma by the
Royal Greenwich Observatory in the Spanish Observatorio del Roque de
Los Muchachos of the Instituto de Astrof\'{\i}sica de Canarias.

\newpage

\figcaption[histo.eps]{Star counts {\it vs.} equivalent widths. 
\label{Fig:histo}}

\figcaption[histo_tipo.eps]{Spectral types in the observed sample.
\label{Fig:histo_tipo}}

\figcaption[histo_tipo_sg.eps]{Spectral types for the supergiants. 
\label{Fig:histo_tipo_sg}}

\figcaption[histo_dist.eps]{Distances. \label{Fig:histo_dist}}

\newpage

\begin{deluxetable}{rrrrrrrrrrrrrrr}

\tablecolumns{15}
\tablewidth{0pt}
\footnotesize
\tablecaption{Coordinates of the TMGS sample.
\label{Tabla}}
\tablehead{\colhead{} &
\multicolumn{13}{c}{Right ascension ($^{\rm h}$ $^{\rm m}$ $^{\rm s}$) and 
Declination (\arcdeg  \arcmin  \arcsec)\ \ \ \  (J2000.0)} & 
\colhead{} }

\startdata
18 30 28.2 & --5 12 34 & S && 18 35 36.0 & --5 03 57 &   &
& 18 39 36.1 & --5 16 48 & S && 18 43 25.9 & --5 11 33 \nl

18 30 52.9 & --5 08 24 &   && 18 35 45.4 & --5 20 15 & S &
& 18 39 49.8 & --5 18 20 &   && 18 44 26.2 & --5 14 49 & S \nl

18 31 57.2 & --5 13 05 & S && 18 36 23.0 & --5 07 04 & S &
& 18 39 58.2 & --5 16 45 & S && 18 44 47.1 & --5 14 51 & S \nl

18 32 04.3 & --5 13 31 & S && 18 37 17.7 & --5 16 12 & S &
& 18 40 01.0 & --5 14 11 &   && 18 45 08.7 & --5 12 34 & S \nl

18 32 17.6 & --5 12 34 & S && 18 37 26.0 & --5 05 08 & S &
& 18 40 01.7 & --5 13 01 &   && 18 45 12.4 & --5 16 50 \nl

18 32 21.3 & --5 14 51 & S && 18 37 45.9 & --5 20 29 & S &
& 18 40 49.0 & --5 05 23 &   && 18 45 34.2 & --5 12 33 \nl

18 32 29.0 & --5 16 50 & S && 18 37 53.2 & --5 04 32 &   &
& 18 40 49.1 & --5 05 28 &   && 18 45 40.0 & --5 07 00 & S \nl

18 32 33.2 & --5 16 49 &   && 18 37 54.5 & --5 14 48 & S &
& 18 41 21.2 & --5 16 17 & S && 18 45 41.9 & --5 14 48 \nl

18 32 34.4 & --5 14 49 & S && 18 38 05.7 & --5 19 37 & S &
& 18 41 36.0 & --5 13 24 & S && 18 45 43.9 & --5 15 21 & S \nl

18 33 05.3 & --5 17 40 & S && 18 38 23.8 & --5 15 42 & S &
& 18 42 03.6 & --5 04 54 &   && 18 45 47.1 & --5 17 45 & S \nl

18 33 06.9 & --5 09 47 &   && 18 38 39.2 & --5 12 22 & S &
& 18 42 14.3 & --5 11 30 & S && 18 45 53.6 & --5 21 05 \nl

18 34 35.2 & --5 11 56 & S && 18 38 50.4 & --5 12 31 & S &
& 18 42 17.8 & --5 13 06 &   && 18 51 37.1 & --5 20 11 \nl

18 34 37.3 & --5 15 05 &   && 18 39 05.4 & --5 12 28 & S &
& 18 42 35.1 & --5 16 50 &   && 18 54 39.8 & --5 11 43 & S \nl

18 35 12.5 & --5 17 44 &   && 18 39 28.2 & --5 14 47 & S &
& 18 42 50.2 & --5 18 23 & S \nl

18 35 35.3 & --5 04 01 & S && 18 39 32.9 & --5 15 19 &   &
& 18 43 15.5 & --5 17 48 & S \nl
\enddata
\tablecomments{S: source has been classified as a supergiant.}
\end{deluxetable}

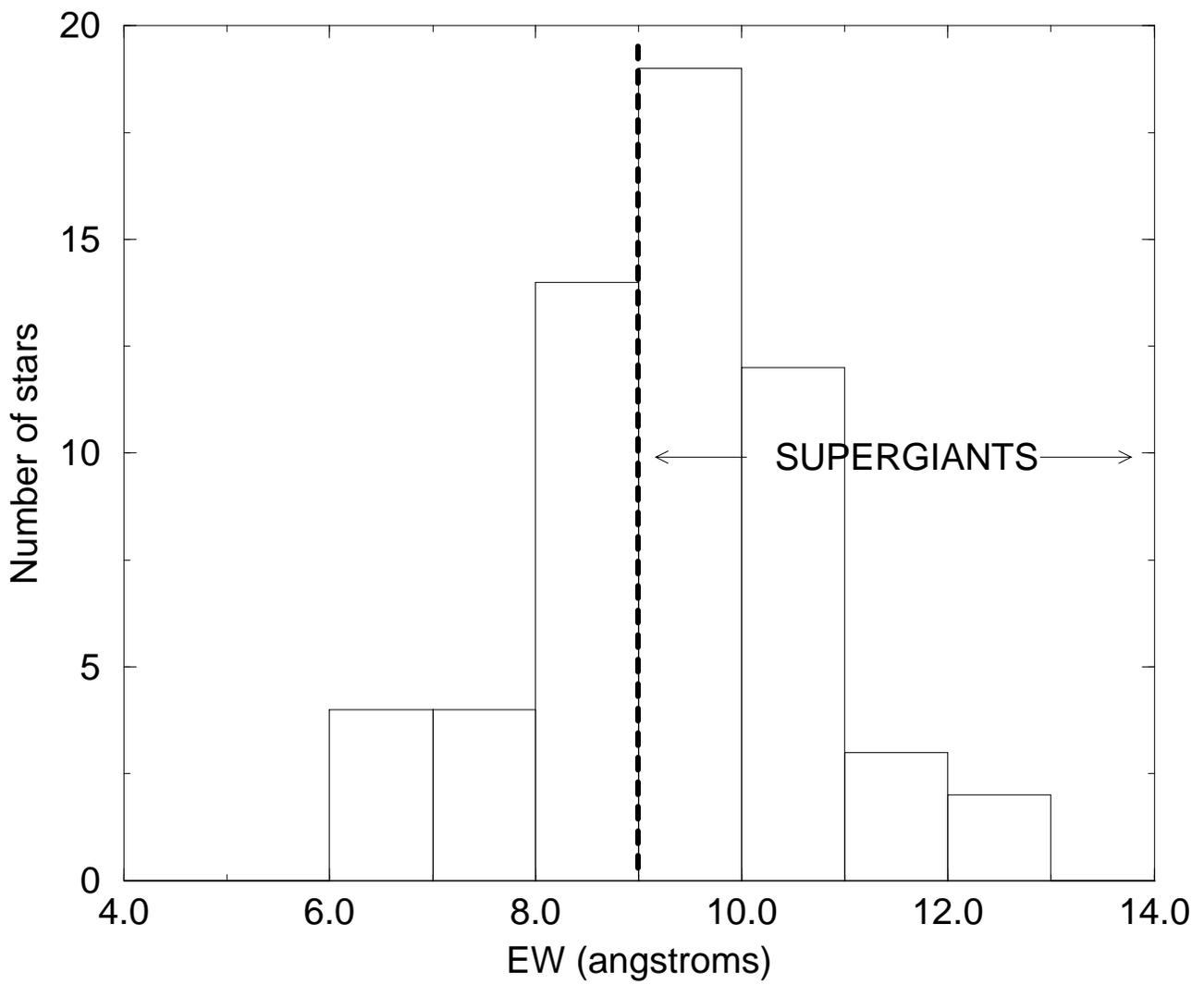
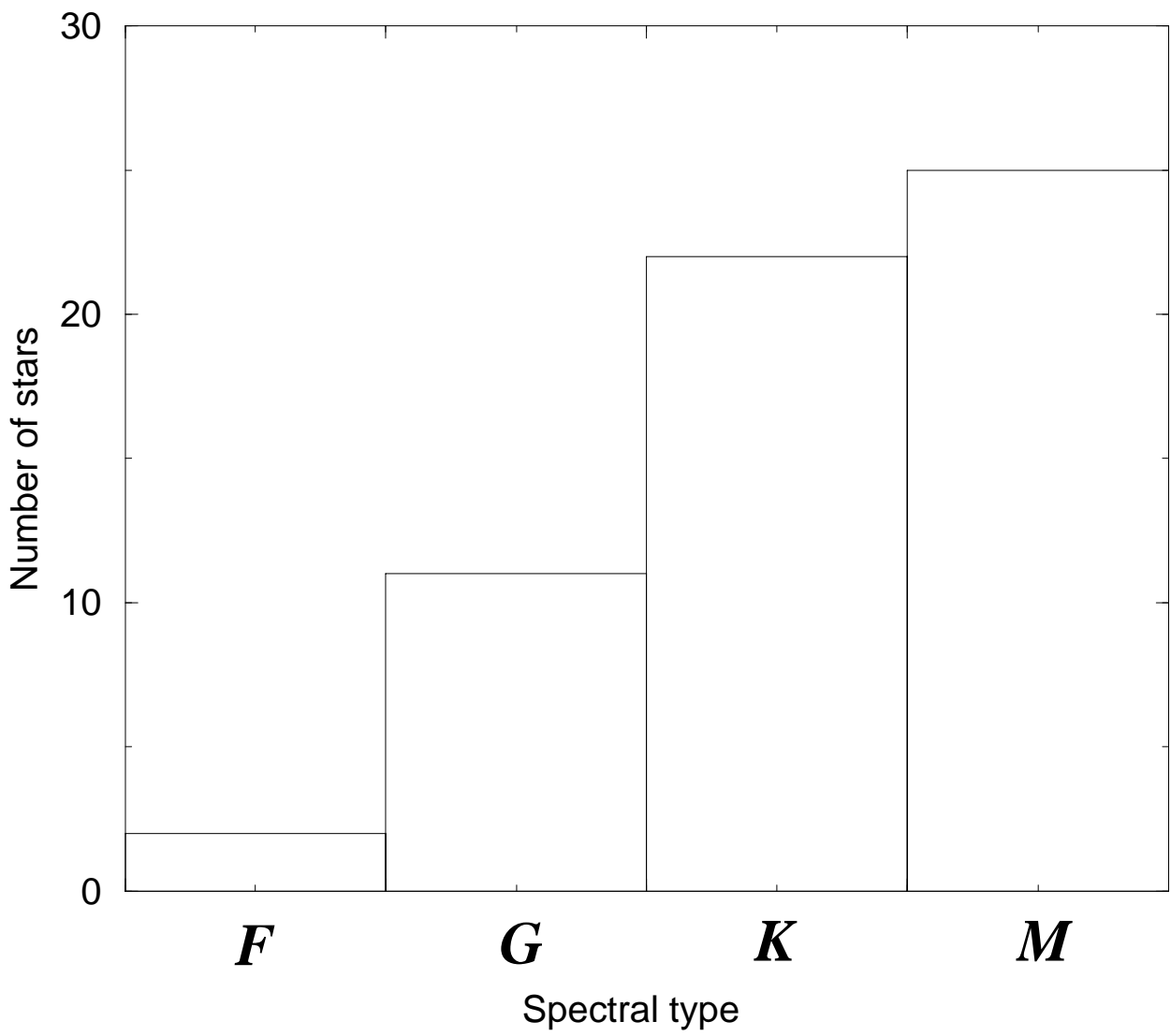
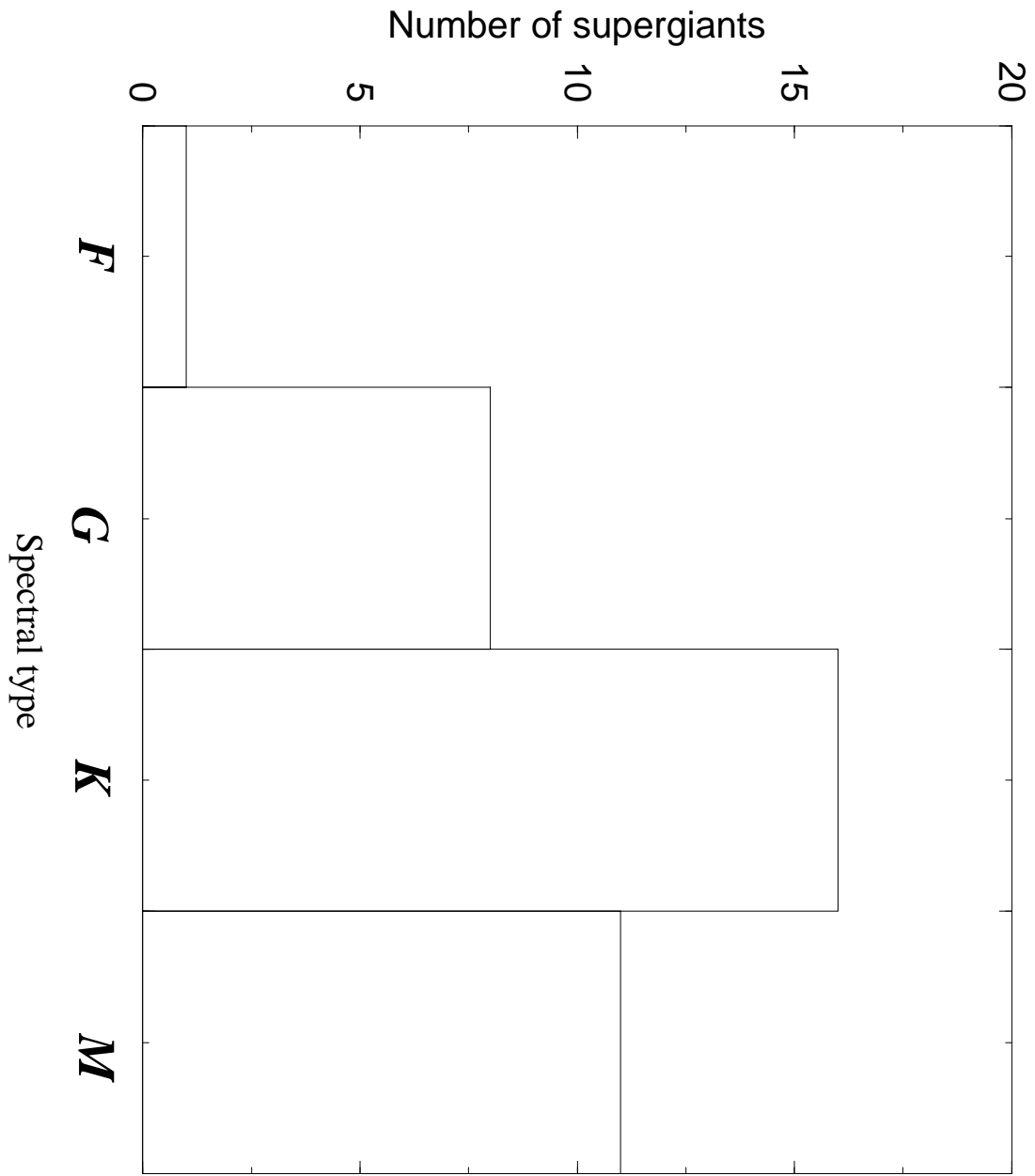
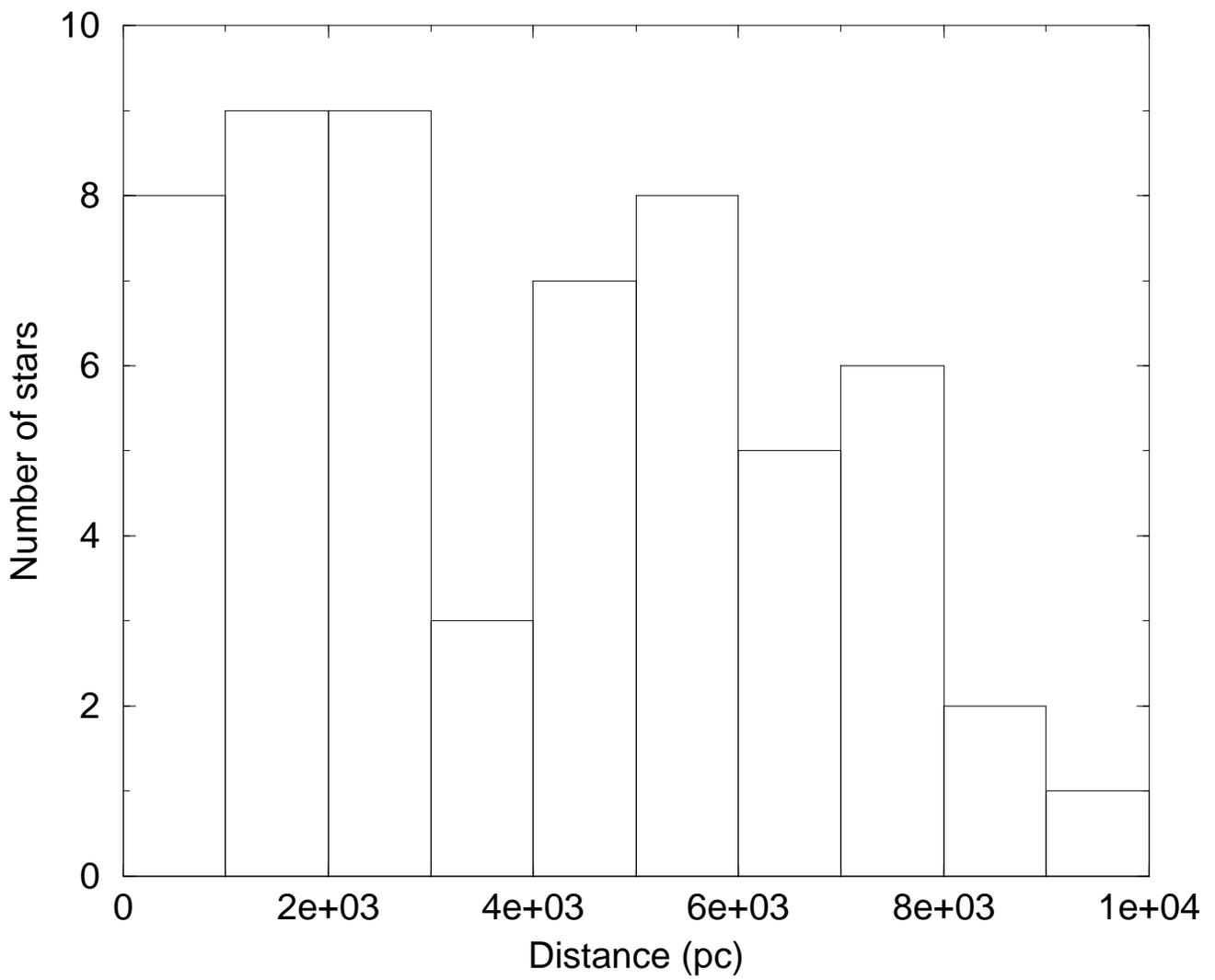
\end{document}